\documentclass[12pt]{article}
\usepackage{amssymb}
\usepackage{amsmath, amsthm}
\usepackage{graphicx}

\newcommand{\be}{\begin{equation}}
\newcommand{\ee}{\end{equation}}

\newcommand{\ba}{\begin{eqnarray}}
\newcommand{\ea}{\end{eqnarray}}
\begin{document}
\def\theequation{\arabic{section}.\arabic{equation}}
\begin{titlepage}
\title{An elementary  derivation of the Deutsch-Jozsa algorithm
}
\author{ Neda Amin and Patrick Labelle\\ 
\\
{\small \it Physics Department, Bishop's University}\\
{\small \it Sherbrooke, Qu\'{e}bec, Canada J1M~0C8}
}
\date{} \maketitle
\thispagestyle{empty}
\vspace*{1truecm}
\begin{abstract}
Quantum computing takes fully advantage of the superposition principle to increase greatly (even exponentially) the speed of calculations, relative to the classical approach. 
The Deutsch-Jozsa algorithm is the simplest quantum algorithm illustrating  this power. Unfortunately,  the standard derivation involves several ingenious steps which usually leave students feeling that they could never have figured out the algorithm by themselves. We present here a different formulation of the problem which allows students to derive the algorithm using only basic knowledge of quantum mechanics and linear algebra.  
\end{abstract} \vspace*{1truecm}
\setcounter{page}{1}
\end{titlepage}

\tableofcontents
\clearpage

\def\theequation{\arabic{section}.\arabic{equation}}


\section{Introduction}
\setcounter{equation}{0}
\setcounter{page}{2}

Some of the most exciting  applications of modern quantum mechanics are  in the field of quantum computing. As such it would be worthwhile to introduce the topic in undergraduate quantum mechanics classes, even only briefly. Although the basic idea of quantum computing is simple,  to harness the power of the superposition principle to speed up calculations,   the actual design of quantum algorithms  is far from intuitive.

The simplest and among the most well-known  is the Deutsch-Jozsa algorithm which we will describe in details below (introduced in Refs.\cite{D1} and \cite{D2} and subsequently improved in Ref.\cite{D3}) . The basic idea of the algorithm is to show how, for a  specific type of operator,   treating the system quantum-mechanically and making  a  clever choice for the input quantum state  allows one to gain information  in fewer operations than would be possible with any classical system. This is of course the essence of the power of quantum computing but the beauty of the Deutsch-Jozsa algorithm is that it exhibits this power in the  simple setting  of a two dimensional Hilbert space and without requiring any other mathematical concept than basic linear algebra (as opposed to Shor's algorithm, for example,  which involves quantum Fourier transforms).  As such it is the ideal example of quantum computing for an introductory quantum mechanics class.

It is fairly straightforward to verify that the particular quantum superposition provided by the algorithm  does indeed allow to gain information in fewer steps than would be possible classically  and checking this, in itself,   can be assigned as a homework problem.   However, it is not obvious how one could have seen that this input quantum state   is the right choice to start with. In other words, it is easy to check that the solution works but it is hard to see how one  could have figured it out  in the first place. After checking that the algorithm works, students therefore often ask: How did anyone ever come up with this solution? 
Of course, the discoverers of the algorithm had a lot of experience in quantum mechanics and are very clever,  which is why they were able to find the solution. Indeed, even the standard presentation of the solution (presented below) involves several steps which are nontrivial and which usually leave the students feeling that they could never have thought of that by themselves.

From a  pedagogical point of view it would be more interesting to have the students figure out the algorithm by themselves as opposed to simply applying it. The goal of this short article is to show how the  problem can be posed in such a way that it can be solved  by students taking   an introductory quantum mechanics class  and   using  none of the clever tricks involved in the standard presentation but only basic linear algebra.

In the first section we review  the problem to be solved. In the following section we  give the standard presentation of the Deutsch-Jozsa algorithm.  In the last section we present a way to state the problem such that the algorithm can be ``discovered" by the students using only simple algebra and a basic understanding of quantum mechanics. The solution is then	 presented.

\section{The problem to be solved}

Consider for now a classical bit whose  states we will denoted by $0$ or $1$. In this section we consider only classical systems. Consider a function which acts on a single bit\cite{foot1}. Let's  define a {\it constant} function one which always returns the same value no matter what the input is. Clearly, there are two such constant functions: the one which always returns 0, which we will denote $C_I$, and the one which always returns 1, denoted by $C_{II}$.

In other words, we have that the action of $C_I$ on a bit is given by
\ba
C_I(0) &=& 0, \\
C_I(1) &=& 0 \ea
whereas
\ba
C_{II}(0) &=& 1, \\
C_{II}(1) &=& 1 .\ea

Now we consider the so-called {\it balanced} functions, which are balanced in the sense that they may return 0 or 1 depending on their input. There are also two possible balanced functions. We will call the first one $B_I$,  with action

\ba
B_I(0) &=& 1, \\
B_I(1) &=& 0 .\ea
The second balanced function, $B_{II}$ is simply the identity operator:
\ba
B_{II}(0) &=& 0, \\
B_{II}(1) &=& 1. \ea

Clearly, the balanced operators are invertible; given one of the two operator and a certain output one can determine uniquely what the input was. This is obviously not the case for the constant functions.

Another operation that is required before we can proceed is the {\it exclusive OR} (XOR) operation $\oplus$ which is simply the addition of two bits (in base 2):
\be 0 \oplus 0 = 1 \oplus 1 = 0~~~~~~~~~~~~~~~0 \oplus 1 = 1 \oplus 0 = 1. \ee
Another way to think of XOR is as a parity operation. Indeed, if one assigns a positive parity to 0 and a negative parity to 1, the operation $ a \oplus b$ essentially computes the parity of the product $ab$.

Consider now an  operator that  takes two bits $x$ and $y$  as  an input.   The first bit, $x$, is used as a control bit and is therefore left unchanged.  The second bit is replaced by
\be y \rightarrow f(x) \oplus y \ee
where the function $f$ is one of the four functions described above.  
We will represent such an operator  by the symbol ${\cal F}_f$  where $ f$ is either $B_I$, $B_{II}$, $C_I$ or $C_{II}$. We can therefore represent the operation of ${\cal F}_f$  on two bits  by
\be {\cal{F}}_{f}~(x,y) = (x,f(x) \oplus y). \ee



\bigskip
\bigskip

Note that the operators ${\cal F}_f$  are invertible; they are  actually their own inverse. Indeed, if we apply any which one  twice in a row we get
\be {\cal F} ({\cal F}(x,y)) = {\cal F} (x, f(x) \oplus y) = (x, f(x) \oplus f(x)  \oplus y ) = (x,y) \ee
where we have used that  for any bit x we have $f(x) \oplus f(x)=0$ and therefore $f(x) \oplus f(x) \oplus y = y$.

 To consider a specific example, pick an operator with  the function  $f $ being  the first balanced function $B_I$. If we feed to this operator the two classical bits $x=1,y=1$, the output will then be 
\ba
{\cal F}_{B_I}(1,1)  = (1, 1  \oplus B_I(1)) = (1, 1 \oplus 0) = (1,1) \ea

Now the problem to be solved can finally be stated. 
 Imagine being provided one of the operators ${\cal F}_f$  but not being told which of the four functions it uses to calculate the output. Such an operator with $f$ being unspecified will be called an {\it{oracle}} and will be represented by ${\cal O}_f$ with the understanding that $f$ is unknown.  You may only feed it pairs of bits and examine the result coming out. The question to answer is the following: {\it what is the minimum number of times the oracle must be run  in order to determine with certainty if the function it uses is balanced or constant?}
 
 Of course, another obvious question would be to ask how many trials would be required in order to tell which of the four functions is being used, and we will get back to this point below,  but this is not the question addressed by the Deutsch-Jozsa algorithm.
 
 It is not hard to convince oneself that two trials are required in order to being able to tell if the function is constant or balanced. This will be obvious if we get back to our above explicit example where   the input was  $(1,1)$ and the output was also $(1,1)$. In our example we computed the output using $f= B_I$ but let's pretend that we are only given the input and the output and that we don't know what the function $f$ is.  What can we tell about $f$ from this single run?   The first bit is always left unchanged so we can't learn anything from it. The only information we have is from the second bit. We know that  the  function used by the oracle satisfies  
\be f(1) + 1 = 1 \ee
which implies 
\be f(1) =0.\ee This is the only information provided by the above test run. But there are two functions having this property: $B_I$ and $C_I$.  Therefore, given only that the input was (1,1) and the output (1,1), we can only say that the function is either $B_I$ or $C_I$. To determine which is actually used, we would have to run a second trial. For example, feeding the state $(0,1$) will produce the output \cite{foot2} $(0,0)$ if the function is $B_I$ as in our above example but would have produced the output $(0,1)$ if the function  had been $C_I$.

It is clear that running the oracle only once for some input $(x,y)$ gives either the information $f(x) =1$ or $f(x)=0$. In either case, there are always two possible functions to choose from, one balanced and one constant. The oracle must be run a second time to pick which one.

The above conclusion is inescapable if the bits are classical. The magic of quantum computing is that if we allow the oracle to act on quantum bits (linear superposition of classical bits) it is possible to determine if the function is balanced or constant with a {\it single} run of the oracle (but one can't determine {\it which} of the two balanced or constant functions is being used). This can be done if one is clever about the choice of the two qubits being fed to the oracle. The determination of this clever choice of the input quantum state is what the Deutsch-Jozsa algorithm accomplishes. In the following section we review  the standard presentation of the algorithm.

\section{The Deutsch-Jozsa algorithm: standard presentation}
Te Deutsch-Jozsa algorithm provides a clever choice of the two qubits   to input in the oracle in order to determine if the function is constant or balanced with a single run. Instead of simply stating the answer  we will try to provide the motivation for this choice.

This time the oracle takes as input a tensor product of two qubits and outputs another tensor product of two qubits.
The effect of the oracle on a tensor product of the basis states  is  the following:
\ba
{\cal{O}}_{f} (|0\rangle \otimes  |0\rangle) &=& |0 \rangle \otimes |f(0) \oplus 0 \rangle, \nonumber \\
{\cal{O}}_{f} (|0\rangle \otimes  |1\rangle) &=& |0 \rangle \otimes |f(0) \oplus 1 \rangle, \nonumber \\
{\cal{O}}_{f} (|1\rangle \otimes  |0\rangle) &=& |1 \rangle \otimes |f(1) \oplus 0 \rangle, \nonumber \\
{\cal{O}}_{f} (|1\rangle \otimes  |1\rangle) &=& |1 \rangle \otimes |f(1) \oplus 1 \rangle .
\ea

A general input state may be written in the form $ |a \rangle \otimes |b \rangle $ with
\ba |a\rangle &=& a_1 |0\rangle + a_2 |1 \rangle \nonumber  \\ |b \rangle &=& b_1 |0 \rangle + b_2 | 1\rangle . \label{qub}
\ea 

Since the oracle is a linear operator, we immediately obtain the result for the application to an arbitrary state to be (using Eq.(\ref{qub}) for the two kets)
\ba
 {\cal{O}}_{f} (|a\rangle \otimes  |b\rangle)  &=& a_1 b_1 ~{\cal{O}}_{f} (|0\rangle \otimes  |0\rangle) +a_1 b_2 ~{\cal{O}}_{f} (|0\rangle \otimes  |1\rangle) \nonumber \\
 &+& a_2 b_1~ {\cal{O}}_{f} (|1\rangle \otimes  |0\rangle) + a_2 b_2 ~ {\cal{O}}_{f} (|1\rangle \otimes  |1\rangle)
 \nonumber \\
 &=& a_1 b_1 |0 \rangle \otimes  |f(0) \oplus 0\rangle +  a_1 b_2 |0\rangle \otimes  |f(0) \oplus 1\rangle \nonumber  \\ &+&  a_2 b_1 |1\rangle  \otimes  |f(1) \oplus 0\rangle + a_2 b_2 |1\rangle  \otimes  |f(1) \oplus 1\rangle. \ea
 Note that  in general this is not a tensor product of the control qubit $|a\rangle$ times some other ket, {\it {i.e.}} it cannot be written  in the form $|a \rangle  \otimes  |c \rangle$ for some ket $|c \rangle$. We therefore see that  in general the control qubit $|a \rangle$  gets entangled with the second input qubit.

The first trick  is to observe  that one way to distinguish balanced and constant functions is to compute $f(0) \oplus f(1) $. Indeed,  the result is
\be f(0) \oplus f(1) = 0 \label{constant}\ee
  if the function is constant and
\be f(0) \oplus f(1) = 1 \label{balanced} 
\ee  if the function is balanced,
as can easily be verified. 
 
At first, it seems as if using this will require running the oracle twice since the function apparently has to be applied twice (once on $0$ and once on $1$).  But this is not necessary if  one works with quantum bits at the condition that the function appears through a phase, as we will now show.

 Schematically, we are trying to have a situation where a relative phase of the desired form will be generated between the two input bits  $|0\rangle$ and $|1 \rangle $. 
 In other words we would like to find a qubit $|b\rangle$ which has the properties
 \ba  {\cal{O}}_{f} (|0\rangle \otimes  |b\rangle) &=& 
  (-1)^{f(0)} |0 \rangle \otimes |b \rangle  \label{propa} \ea
  and
  \ba
{\cal{O}}_{f} (|1\rangle \otimes  |b\rangle) &=& 
  (-1)^{f(1)} |1 \rangle \otimes |b \rangle . \label{propb}
\ea

As we will see below, it is not difficult to find a ket $|b \rangle$ that has those properties.  But before presenting the (almost obvious) answer, let us first discuss why having such a state allows one to distinguish constant from balanced functions  in only one step.

Assuming that we have a state $|b \rangle$ satisfying the above properties, applying the oracle to the product $|a \rangle \otimes |b \rangle$ where $|a \rangle$ is an arbitrary qubit will give 
\ba 
{\cal{O}}_{f} (|a\rangle  \otimes  |b\rangle)  &=& a_1 ~
{\cal{O}}_{f} (|0\rangle \otimes  |b\rangle) + a_2 {\cal{O}}_{f} (|1 \rangle \otimes  |b\rangle) \\ &=&  (-1)^{f(0)}~ a_1 |0\rangle  \otimes  |b \rangle + (-1)^{f(1)}~ a_2 |1\rangle  \otimes  |b \rangle \\ &=& (-1)^{f(0)}  \biggl( a_1 |0\rangle  \otimes  |b \rangle + (-1)^{-f(0)+f(1)}~ a_2 |1\rangle  \otimes  |b \rangle \biggr)\\ &=& (-1)^{f(0)} \biggl(a_1 |0\rangle  + (-1)^{f(0)+f(1)} a_2 |1 \rangle  \biggr) |b \rangle 
 \ea 
 where  have used $(-1)^{-f(0)} = (-1)^{f(0)}$. 
  As usual, the overall phase is unimportant.

Therefore, using that $f(0) \oplus f(1) = 1$ if the function is balanced (see Eq.(\ref{balanced})), the final result is simply (after discarding the overall phase)
\ba
{\cal{O}}_{f} (|a\rangle  \otimes  |b\rangle) &=& \biggl( a_1 |0 \rangle - a_2 |1 \rangle \biggr)  \otimes |b \rangle 
\nonumber \\&=& a_1 |0 \rangle \otimes |b \rangle - a_2 |1 \rangle \otimes |b \rangle ~~~\text{(for a balanced function).} 
\ea
 and if the function is constant, using $f(0) + f(1) =0$,  we  simply get 
\ba
{\cal{O}}_{f} (|a\rangle  \otimes  |b\rangle) &=& |a \rangle  \otimes |b \rangle \nonumber 
\\ &=& a_1 |0 \rangle \otimes |b \rangle + a_2 |1 \rangle \otimes |b \rangle~~~\text{(for a constant function.)} \ea

If we impose that the two possible outcomes are orthogonal so that they can be distinguished, we find that we must have $|a_1|^2 - |a_2|^2 = 0 $
which implies that $a_1$ and $a_2$ may only differ by a phase: 
\be a_2 = e^{i \theta} a_1 \ee Therefore,  the normalized control qubit must be 
\be |a \rangle = \frac{ |0 \rangle  + e^{i \theta} |1 \rangle }{\sqrt{2}}  \label{sol} \ee
and the input state that must be fed to the oracle is
\be
| a \rangle \otimes | b \rangle  = \frac{ |0 \rangle  + e^{i \theta} |1 \rangle }{\sqrt{2}} \otimes |b \rangle \ee
with the ket $|b \rangle$ having the properties (\ref{propa}) and (\ref{propb}). 
    
We have accomplished the goal we had set for ourselves: we have found an input state such that running the oracle only once will determine if the function is constant or balanced.  Indeed, if the we use the above state as input,  we simply have to project the output on the bra $\langle a | \otimes \langle  b|$. If the result is one, the function used by the oracle is constant. If the result is zero, the function is balanced.

All this  relies of course on finding a state  $|b \rangle $ satisfying the properties (\ref{propa}) and (\ref{propb}). It is not difficult to guess what the answer is:
\be |b \rangle = \frac{ |0 \rangle  - |1 \rangle }{\sqrt{2}} . \label{yy} \ee
Consider the case $f(0) =1$ (so the function is either $B_I$ or $C_{II}$ ). Then we have
\ba {\cal{O}}_{f} ( |0 \rangle \otimes | b \rangle ) &=& \frac{{\cal{O}}_{f} (| 0 \rangle \otimes |0 \rangle)}{\sqrt{2}} -\frac{ {\cal{O}}_{f} (
|0 \rangle \otimes |1 \rangle)}{\sqrt{2}} \nonumber \\
&=&  \frac{|0 \rangle \otimes | f(0) \oplus 0 \rangle}{\sqrt{2}} -  \frac{|0 \rangle \otimes | f(0) \oplus 1 \rangle}{\sqrt{2}} 
\nonumber \\
&=& \frac{|0 \rangle \otimes | 1 \rangle}{\sqrt{2}} - \frac{|0 \rangle \otimes |0 \rangle}{\sqrt{2}} \nonumber \\
&=& - |0 \rangle \otimes |b \rangle \ea
which we may write as $ (-1)^{f(0)} |0 \rangle \otimes |b \rangle$.

It is easy to verify that if $f(0) =0$ (so the function is either $C_I$ or $B_{II}$), the state $|0 \rangle \otimes |b \rangle $ is  left unchanged by applying 	${\cal{O}}_{f}$ so that we may still write  the result as  $ (-1)^{f(0)} |0 \rangle \otimes |b \rangle$. The conclusion is therefore that for any of our four operators, we obtain
\be
{\cal{O}}_{f} ( |0 \rangle \otimes | b \rangle ) = (-1)^{f(0)} |0 \rangle \otimes |b \rangle
\ee
with $|b \rangle$ chosen as in Eq. (\ref{yy}).

It is easy to check also that for any operator, one  finds
\be
{\cal{O}}_{f} ( |1 \rangle \otimes | b \rangle ) = (-1)^{f(1)} |1 \rangle \otimes |b \rangle .
\ee

 So the state (\ref{yy}) satisfies the relations (\ref{propa}) and (\ref{propb}). 

Our final result is therefore that the input we must feed to the oracle is 
\be
\biggl( \frac{ |0 \rangle  + e^{i \theta} |1 \rangle }{\sqrt{2}}  \biggr) 
      \otimes \biggl( \frac{ |0 \rangle  - |1 \rangle }{\sqrt{2}} \biggr) . \label{finalresult} \ee
 With this choice of input, the outputs corresponding to an oracle using a balanced function will be orthogonal to the output corresponding to an oracle using a constant function.  As explained above, one can determine if the function is constant or balanced by projecting on the bra
 $\langle a | \otimes \langle  b|$.
 
\section{An elementary derivation}
The presentation given  in the previous section emphasizes the key ideas on which the algorithm is based but students (and non-students alike!) may feel that it relies on too many clever guesses. 
We will present now the problem in such a way that it can be solved ``mechanically" using only basic linear algebra  and none of the clever tricks needed in the previous derivation. It is of course more  algebra intensive and may be considered less satisfying from a conceptual point of view by some. But it has the pedagogical advantage that it can be solved  by students with only a basic knowledge of quantum mechanics who will therefore have the satisfaction of having ``rediscovered" the algorithm by themselves.

 We will represent the tensor products of two qubits as four component column vectors using the convention
 \ba |0\rangle  \otimes |0\rangle &=& \left( \begin{array}{c} 1 \\0 \\0 \\0 \end{array} \right), ~~~~~~~~~
 |0\rangle \otimes  |1\rangle =  \left( \begin{array}{c} 0 \\1 \\0 \\0 \end{array} \right), \\ 
 |1\rangle \otimes  |0\rangle  &=&  \left( \begin{array}{c} 0 \\0 \\1 \\0 \end{array} \right), ~~~~~~~~~
 |1\rangle \otimes |1\rangle =  \left( \begin{array}{c} 0 \\0 \\0 \\1 \end{array} \right).
 \ea

The first step is to represent the four possible operators ${\cal{F}}_{f}$ as four by four matrices.
It is now easy to construct explicit representations of those operators since we know how they act on each of the four basis states.  For example,
\be  {\cal{F}}_{C_{II}} (|0 \rangle \otimes  |0 \rangle) = |0 \rangle \otimes |C_{II}(0) \oplus 0 \rangle = |0\rangle \otimes  |1 \rangle \ee
and so on. One finds
\ba {\cal{F}}_{C_I} &=& 
\left(
\begin{array}{cccc}
1 & 0 & 0 & 0
\\
0 & 1 & 0 & 0
\\
0 & 0 & 1 & 0
\\
0 & 0 &0 &1
\end{array}
\right),  ~~~~~~~~~ {\cal{F}}_{C_{II}} = 
\left(
\begin{array}{cccc}
0 & 1 & 0 & 0
\\
1 & 0 & 0 & 0
\\
0 & 0 & 0 & 1
\\
0 & 0 &1 &0
\end{array}
\right), \\
 {\cal{F}}_{B_I} &=& 
\left(
\begin{array}{cccc}
0 & 1 & 0 & 0
\\
1 & 0 & 0 & 0
\\
0 & 0 & 1 & 0
\\
0 & 0 &0 &1
\end{array}
\right), ~~~~~~~~~ {\cal{F}}_{B_{II}} = 
\left(
\begin{array}{cccc}
1 & 0 & 0 & 0
\\
0 & 1 & 0 & 0
\\
0 & 0 & 0 & 1
\\
0 & 0 &1 &0
\end{array}
\right) .
\ea

Now we are ready to state the problem  in such a way that students in an introductory quantum mechanics class could solve without any subtle trick,  only simple algebra,  and therefore recover the solution provided by the Deutsch-Jozsa algorithm.

Let's consider an arbitrary input state
\be
  |\psi \rangle = \left( \begin{array}{c} c_1 \\c_2  \\c_3 \\c_4 \end{array} \right) 
\ee
with   
\be |c_1|^2 + |c_2|^2 + |c_3|^2 + |c_4|^2 = 1 . \label{normal} \ee

The goal is to choose the coefficients in such a way that, given an oracle, we will be able to tell with only one application whether the function used by the operator is balanced or constant. The way to obtain this is clear: we must choose the coefficients of our input state such that if we run it through each of the four operators,  the output of each balanced operator must be orthogonal to the output of both constant operators.  This gives four conditions which we can schematically write as
\ba 
\langle {\cal{F}}_{C_I} \psi | {\cal O}_{B_I} \psi \rangle &=& 0, \label{first} \\ 
\langle {\cal{F}}_{C_I} \psi | {\cal O}_{B_{II}} \psi \rangle &=& 0, \label{second} \\
\langle {\cal{F}}_{C_{II}} \psi | {\cal O}_{B_I} \psi \rangle &=& 0, \label{third} \\ 
\langle {\cal{F}}_{C_{II}} \psi | {\cal O}_{B_{II}} \psi \rangle &=&0 . \label{fourth} 
\ea
 This is all there is to it. If one can find coefficients satisfying these four conditions (and of course we know that there is at least one solution), the problem has been solved.
 
 Equation (\ref{first}) yields the following condition:
 \ba
 &&c_2^* c_1 + c_1^* c_2 + |c_3|^2 + |c_4|^2 \nonumber \\ &&~= 2 \text{Re} \, (c_1 c_2^*) + |c_3|^2 + |c_4|^2 = 0 \label{fcond} \ea
 whereas equation (\ref{second}) gives
 \be 2 \text{Re}(c_3^* c_4) + |c_1|^2 + |c_2|^2 = 0 . \label{scond} \ee
 The third and fourth equations, (\ref{third}) and (\ref{fourth}), end up being exactly the same as the two above. So we need to solve those two equations with the constraint that the state be normalized, equation(\ref{normal}).  To simplify the problem, one may at first try to see if  there  exists a solution with all coefficients real. Our set of equations is therefore 
 \ba 
 2 c_1 c_2 + c_3^2 + c_4^2 &=& 0, \label{one} \\
 2 c_3 c_4 + c_1^2 +c_2^2 &=& 0, \label{two} \\
 c_1^2 + c_2^2 + c_3^2 + c_4^2 &=& 1 .\label{three} \ea
 If we isolate $c_3^2 + c_4^2 $ from equation(\ref{one}) and use this in equation (\ref{three}), we obtain
 \be c_1^2 + c_2^2 - 2 c_1 c_2 = 1 \Rightarrow (c_1 - c_2)^2 = 1 \ee
 which gives
 \be c_1 = c_2 \pm 1 . \ee
 Similarly, if we isolate $c_1^2 + c_2^2 $ from (\ref{two}) and put this in equation (\ref{three}), we get
 \be c_3 = c_4 \pm 1 . \label{cthree} \ee
 
 \subsection{First case: $c_1 = c_2 +1 $ and $c_3 = c_4 +1 $}
 Let's first pick the cases 
 \ba c_1 &=& c_2+1, \label{yaya} \\ c_3 &=& c_4 + 1 . \label{yoyo}  \ea If we plug those values in the normalization condition (\ref{three}), we get a quadratic formula for $c_2$ in terms of $c_4$:
 \be c_2 = \frac{-1 \pm \sqrt{-4 c_4^2 -4 c_4 -1 }} { 2} . \label{quad}\ee
 If we impose that $c_2$ be real, the discriminant must be larger of equal to zero:
 \be 
 -4 c_46^2 -4 c_4 -1  \geq 0 \ee
 but it's trivial to show that  the only solution is for a strict equality (in other words, this describes a parabola with its apex  on the horizontal axis)  which is solved by
 \be c_4 = - \frac{1}{2} . \ee
 Putting this back in  equations (\ref{cthree}), (\ref{yaya}), (\ref{yoyo}) and (\ref{quad}), we final get 
 \be c_1=-  c_2 = c_3 = -c_4 = \frac{1}{2} \ee
 as our first solution. Written as a tensor product of two kets, this corresponds to 
 \be 
 \frac{1}{2} \biggl( |0\rangle \otimes |0 \rangle - |0 \rangle \otimes |1 \rangle + |1 \rangle \otimes |0 \rangle - |1 \rangle \otimes |1 \rangle \biggr) = \bigl( \frac{|0 \rangle + |1 \rangle)}{\sqrt{2}} \bigr)~~\otimes~~\bigl( \frac{|0 \rangle - |1 \rangle)}{\sqrt{2}} \bigr)
 \ee
which is indeed a special case of equation(\ref{finalresult}) with $ \theta = 0$.

 \subsection{Second case: $c_1 = c_2 +1 $ and $c_3 = c_4 -1 $}
 In that case, one finds obviously the same quadratic equation as before except that with the opposite sign for the term linear in $c_4$:
 \be c_2 = \frac{-1 \pm \sqrt{-4 c_4^2 +4 c_4 -1 }} { 2} .\ee
 Again, imposing $c_2$ to be real leads to solving
 \be
 -4 c_4^2 +4 c_4 -1 \geq 0 \ee
 which, again, turns out to have a unique solution, corresponding to a strict equality. The result this time is $c_4 =  \frac{1}{2}$ which yields 
 \be c_1 = -c_2 = -c_3 = c_4 = \frac{1}{2} \ee
 giving the state 
   \be 
 \frac{1}{2} \biggl( |0\rangle \otimes |0 \rangle - |0 \rangle \otimes  |1 \rangle - |1 \rangle \otimes  |0 \rangle + |1 \rangle \otimes  |1 \rangle \biggr) = \bigl( \frac{|0 \rangle - |1 \rangle)}{\sqrt{2}} \bigr)~~\otimes~~\bigl( \frac{|0 \rangle - |1 \rangle)}{\sqrt{2}} \bigr)
 \ee
 which is indeed the other solution of equation(\ref{finalresult}) with real coefficients, {\i.e.} $\theta = \pi$. 
 
 \subsection{The other two cases}
 We still have to consider the cases $c_1 = c_2 -1 $ and $ c_3 = c_4 \pm 1 $. As one would expect, these two cases give the same two results we obtained above excepted for an overall irrelevant factor of $-1$.
 \section{A Bonus}
 
 One obvious question that comes to mind when learning about the Deutsch-Jozsa algorithm is whether it would be possible to not only determine if the oracle uses a constant or balanced function in only one step but in addition to determine uniquely which of the four functions is used.  Since the standard presentation relies on several clever guesses, one might wonder if more clever tricks could be used to identify uniquely the function. The answer is easy to find using our ``mechanical" approach. The question becomes whether it is possible in addition to the conditions (\ref{first}) - (\ref{fourth}) to fulfill the extra constraints 
 \ba
\langle {\cal{F}}_{C_I} \psi | {\cal{F}}_{C_{II}} \psi \rangle &=& 0 \label{fifth} \\ 
\langle {\cal{F}}_{B_I} \psi | {\cal{F}}_{B_{II}} \psi \rangle &=& 0 \label{sixth} 
 \ea
 which  both give  (going back to complex coefficients for now in order to be more general):
 \ba
 c_1^* c_2 + c_2^* c_1 + c_3^* c_4 + c_4^* c_3 = 0 \ea
 which may be written as
 \be \text{Re}(c_1^* c_2) + \text{Re}(c_3^* c_4) = 0 .\ee
 But this condition is clearly inconsistent with equations (\ref{fcond}) and (\ref{scond}) which combined imply
 
 \be \text{Re}(c_1^* c_2) + \text{Re}(c_3^* c_4) =  - \frac{1}{2} . \ee
 Therefore we see that it would be impossible to determine uniquely the function used by the oracle in a single step.

 \section{Conclusion}
 We have  
  shown  how it is possible to formulate the problem solved by the Deutsch-Jozsa algorithm   in such a way that  the solution can be found using only  basic concepts of quantum mechanics and simple linear algebra. This brings the rediscovery of the algorithm within the reach of students taking an introductory class in quantum mechanics as part of an assignment set, for example.
 



\begin{thebibliography}{5}
\bibitem{foot1} The problem can be generalized to operators having for output a single bit but taking for input $n$ bits. We will not consider this more general case here.
\end{thebibliography}
\end{document}